\documentclass{icst}

\usepackage{moreverb}

\usepackage[breaklinks,colorlinks,bookmarksopen,bookmarksnumbered,linkcolor=ICSTblue,citecolor=blue,urlcolor=ICSTblue]{hyperref}
\usepackage{breakurl}
\usepackage{doi}

\newcommand\BibTeX{{\rmfamily B\kern-.05em \textsc{i\kern-.025em b}\kern-.08em
T\kern-.1667em\lower.7ex\hbox{E}\kern-.125emX}}

\def\volumeyear{2014}

\journalname{XXXXXX}
\articletype{Research Article/Editorial}
 \def\volumeyear{XXXX}
 \def\volumenumber{XX}
  \def\issuemonth{XX-XX}
  \def\issuenumber{X}
  \def\articleid{XX}
  \def\copyrightholder{Author Name\textit{ et al.}, licensed to ICST}
  \copyrightnote{This is an open access article distributed under the terms of the Creative Commons Attribution license (\url{http://creativecommons.org/licenses/by/3.0/}), which permits unlimited use, distribution and reproduction in any medium so long as the original work is properly cited.}
  \def\articledoi{10.4108/XX.X.X.XX}
\received{XXXX}
  \accepted{XXXX}
  \published{XXXX}

\begin{document}

\runningheads{S. Sudha}{A demonstration of the \journalabb\ class file}

\title{Extracting Actionable Knowledge from Domestic Violence Discourses on Social Media}

\author{Sudha Subramani\affil{1}, Manjula {O}'Connor \affil{2}}

\address{\affilnum{1}Centre for Applied Informatics,Victoria University, Melbourne, Australia\\ \affilnum{2} Department of Psychiatry, The University of Melbourne, Melbourne, Australia
 } 

\abstract{Domestic Violence (DV) is considered as big social issue and there exists a strong relationship between DV and health impacts of the public. Existing research studies have focused on social media to track and analyse real world events like emerging trends, natural disasters, user sentiment analysis, political opinions, and health care. However there is less attention given on social welfare issues like DV and its impact on public health. Recently, the victims of DV turned to social media platforms to express their feelings in the form of posts and seek the social and emotional support, for sympathetic encouragement, to show compassion and empathy among public. But, it is difficult to mine the actionable knowledge from large conversational datasets from social media due to the characteristics of high dimensions, short, noisy, huge volume, high velocity, and so on. Hence, this paper will propose a novel framework to model and discover the various themes related to DV from the public domain. The proposed framework would possibly provide unprecedentedly valuable information to the public health researchers, national family health organizations, government and public with data enrichment and consolidation to improve the social welfare of the community. Thus provides actionable knowledge by monitoring and analysing continuous and rich user generated content.
}



\fnotetext[1]{Corresponding author.  Email: \email{sudha.subramani1@live.vu.edu.au}}

\maketitle

\section{Introduction}
Domestic Violence (DV) against women [1] is not a new phenomenon, and it causes serious impacts to {women}'s physical, mental health and well-being.  The existing statistical surveys and reports [2] reveal that violence against women is a global public health issue which affects approximately one in three women universally. The findings of existing statistical content analysis and survey results highlight the following apparent research gaps [3], which are addressed in this paper.
\begin{itemize}

\item First, the existing surveys always rely on a sample of the population and requires manual annotation, much human efforts which is labour-intensive, time consuming and expensive.
\item Next, it requires more safety measures to handle this kind of research as women are reluctant to disclose their sufferings, and there is a need to collect data in a private space and in the absence of male partners. The training of interviewers is also to be done properly and interviewing only one woman in the household to inhibit the sharing of survey content between others.
\item	Finally, the existing studies on this issues have the constraints due to the less availability data and only provides the selected number of health outcomes and making it difficult to determine the co-morbidity level, emotions, opinions and feelings. Women suffering from DV are likely to experience not only mental disorders, but also suffer physically, emotionally and psychologically.
\item Hence there is a need to develop the novel methodology other than statistical surveys to analyse the various themes related to DV: \emph{various abuses [physical, verbal, psychological, emotional and sexual],  mental health problems [depression, anxiety, trauma...] and  physical health disorders [fever, head ache, bleed, abortion…], their emotions and mood swings [sad, happy, relax, stress, worries, sympathetic, regrets…], crisis support and services, legal guidance and so on}.
\end{itemize}

\textbf{Research Aim}

The massive amount of data generated over social media is witnessing a foremost development in the last few centuries. Social media encourages the users for their freedom of thoughts and provides a platform to share their expressions, opinions and random details of their lives with no restrictions in the form of posts and status updates. That play a predominant role to analyse public opinion by aggregating the thoughts of millions of users. Though each tweet or post contains less informational value, the consolidation of millions of messages can generate actionable knowledge and provide valuable insights about the public opinion in general. 

Several studies in different domains: news events, natural disasters, user sentiment analysis, political opinions explored Twitter, a fascinating source of public opinions, over half an billion user generated messages posted every day in order to extract useful information. But, less attention was directed toward studying social welfare topics and health care in social media compared to other topics like marketing products, sentimental analysis of customers and politics. In recent years, this has become the active research area that has drawn huge attention among the research community for information retrieval and to discover the abstract topics that underlies on the large microblogging stream. But, it is challenging for the information retrieval from the large streams of microblogs of its following characteristics:
\begin{itemize}
\item	\emph{Immense scale of volume, fast data arriving rate and their unique characteristics of short length.}
\item	\emph{Large number of spelling and grammatical errors.}
\item	\emph{Use of informal and mixed language.}
\item	\emph{Microblog texts are usually relevant to wide-ranging events in real world events. Some of them might carry interesting and relevant information, whereas others might not.}
\end{itemize}

\section{Contributions}
This section starts with discussing the academic 
contribution about the methodologies used in this paper. Then, we explain certain
practical benefits that are related to the public health sector.

\subsection{Contribution to Knowledge (Academic Contribution)}
The contribution of this paper is to take the plethora of information from the information explosion era of social media on a daily basis and gain in-depth insights of knowledge discovery. This work harvests DV related data, identify hidden patterns, generate models, and finally use that data to improve the quality of care. Collecting the data from Twitter’s data warehouse, and run large calculations over the Twitter dataset easily and efficiently, for turning that data into something actionable. While the existing topic models have focused only on term based approaches to detect topics, they have the difficulties in interpretation and scalability for understanding, organizing and exploring the quality topics from microblogging data stream. Hence this paper uses the pattern mining technique to identify the patterns, then the MapReduce architecture is used on the mined patterns and finally cluster the terms which are more coherent to the topic. The endeavor of this paper is to focus on social media and provide appropriate data analytic methods, in order to improve the classification accuracy and high interpretability of the cluster of topics.

\subsection{Statement of Significance (Practical Contribution)}
Recently victims of DV are turning to social media to share their experiences for social support which has two realms like informational and emotional support. Not only victims, but also friends, family members, care givers, family welfare organization, psychiatrist, and police are increasingly sharing their opinions, feelings, and regrets about the incidents to create social awareness. 

To the best of our knowledge, this would be the first work to use social media to analyse the public health informatics, mine out latent topics and internal semantic structures on the topic of DV. These comprehensive findings would mark an important milestone, not only in the field of research in DV against women, but also in the field of global public health in general. This support the people with data enrichment, consolidation about a specific topic and the entire community can be benefited with the valuable information extracted from the large data.

The public health sector plays a significant role by considering the serious health risks faced by women and their families by this evidence based surveys. This promotes and strengthens the health care systems to women who have experienced violence. This report not only provides the first such summary of acts of DV against women, but also would make the significant advancement for women’s health and rights. Let this report serve as a unified call to action for those working for a world without violence against women.

\section{Literature Survey}

With the increasing popularity of social media, several research studies focused on social media to analyse and predict real world activities like user sentiment analysis [4], opinion mining on political campaigns [5][6], natural disasters  [7], epidemic surveillance [8], event detection [9],  e-healthcare services [64] [65] and so on. As twitter plays a vital role in the life of many people, 500 million tweets are shared per day and 200 billion tweets per year [10]. On the other
hand, some studies dealt with security and privacy issues [66]
[67] [68] [69] [70], as the increasing sophistication of social
media data is posing substantial threats to users privacy.  In contrast to statistical surveys, Twitter becomes a reliable data source to share the opinions and thoughts in online immediately with a status update. Hence, this becomes an efficient platform for researchers to detect real world events in informational retrieval and decision making.
                                                                                                                                                                                                                                                                                                               
Sentiment analysis has been previously studied on different aspects such as blogs and forums and has now been analysed in social media [11]. Liu et al. [12] have done Sentiment analysis by extracting the comments on the attributes and features of a particular aspect and categorizes the comments as positive, negative or neutral. The aspect can be a product, event, person, or topic. 

Opinion mining on political events shows that tweets are effective to describe the opinion about an event. O’Connor et al. [6] and Tumasjan et al. [7] have studied that sentimental analysis of tweets correlated with the voters’ political preferences and closely aligned with the election results. Not only in the field of politics, but also in economics, have public tweets played a major role. Bollen et al. [13][14] analysed that tweet sentiments can be used to predict trends of stock  and it is directly correlated with them. Bruns et al. [15], Burns et al.[16],  Gaffney et al.[17] observed that Twitter is a powerful tool to gather public opinion and create social change. 

Sakaki et al. [7] investigated tweets during natural disasters and shown that it is able to detect earthquakes and send warning alerts to society. They considered each twitter user as a mobile sensor in Japan and the probability of an earthquake is computed using time and geolocation information of the user. Posting time and volume were modelled as exponential distribution to estimate locations of earthquake using kalman and particle filters. Their research further evidenced that earthquake can be sensed earlier than official broadcast.

Culotta et al. [18][19] analysed Twitter to detect influenza epidemic outbreaks that improves speed and cost reduction from traditional methods. Data of the user like gender, age and location can be used to provide more descriptive information about demographic insights compared to search queries. They detected influenza using multiple regression models and Quincey et al. [20] detected swine flu from Twitter using pre-defined keywords and terms co-occurrence method. These methods are analysed by searching the tweets with the keywords and detected anomalous change with the rapid flow in message traffic related to given keywords. The aids of such a method is to collect more focused information from the Twitter stream.

Twitter also becomes the effective medium to research in health care topics and analyse various diseases like cholera [21], cardiac arrest [22], dental pain [23], alcohol use [24], tobacco [25], drug use [26], depression [15], Mood swings [27], and Ebola outbreak [28].  Michael et al. proposed a technique called Ailment Topic Aspect Model [29] [30][31] to monitor the health care of public about the diseases, symptoms and treatments.
 
Conventional methods [32][34] of event detection has been addressed in Topic Detection and Tracking [33] that deals with traditional media sources like news articles, academic papers and so on to find the real world events. Event detection is difficult to perform in Twitter because of its noise and short size.  Bernstein et al. [35] proposed interactive topic browsing system that groups twitter in to topics and used tag cloud for visualization. Sankaranarayanan et al. [36] developed TwitterStand facilitates users to browse news based on geographic preference. This extracts the feature sets, performs online clustering based on term frequencies to find topics.  TwitInfo [37] is a similar approach that performs sentimental analysis to assists users and also calculates peaks from frequency of tweets.

Another approach in twitter trend analysis is bursty topic detection, which are trendier in the time series. Lee et al. [38] developed a sliding window model and it considers the time factor of the term frequency. It calculates the term weights based on the arrival rates in a specified time frame. Burstiness is used to determine how popular a term at the given time interval in the particular event detection.

Topic model is one of the popular research areas to identify semantic relationship from text content. The probabilistic topic model by Blei et al. [39] infers the problem of topic detection as probabilistic distribution. It represents the document as a distribution over topics and a topic as distribution over words. On variation of LDA, online-LDA [41], Dynamic topic models [40], labelled LDA [42] used probabilistic topic models as a baseline model to detect events in twitter streams. However, due to the unique characteristics of twitter streams like short length, noisy, informal texts,  spelling mistakes and fast arrival rate, those methods do not suit well. 
LDA models are used to extract the topics [71]. Various classifiers are trained to predict the model performance [72,73,74,75].

Most of the topics use term based models, which are difficult to predict and challenging to users for better interpretation. This suffers from polysemy and synonyms. Polysemy means, one word defines multiple meanings. For instance, the term Apple defines both fruit and technology. Synonym is about the similar meaning of two different words. For instance, both the bank river and financial organization can be called as bank. Hence patterns are the best practice for topic modeling, as the meaning of the word can be expressed well in patterns with the co-occurrence of neighbour terms. 

Pattern mining is the classical approach that is used in traditional databases. Apriori [43][44] is one of the most popular and important association rule mining algorithms.  Association rule mining discovers all the association rules between the item sets that exceed the values of minimum support and confidence. 

Patterns and Interest are the major factors in data mining applications. Association rule mining and classification extract the patterns and the interest factor defines whether the pattern is useful or not. Data mining could also be viewed as process of converting data into information, the information in to value [57], which is used by users. Mining the actionable patterns, this could be beneficial to the people, organization, and community. Hence, in the social media of big twitter data stream, actionable knowledge can be extracted, by mining the interesting patterns related to the particular issue [58]. 

Recent research studies used some techniques like Twitter Monitor [59], EDCoW [60], HUPC [61], and SFPM [62].  The first two techniques are based on bursty keyword analysis.  The keywords that have higher absolute frequency than usual are used to find more bursty terms. EDCoW method applied wavelet analysis to model the terms based on the frequencies. HUPC and SFPM applied pattern mining process to detect hot topics from Twitter data streams.

\section{Proposed Methodology}
There are a lot of studies in pattern mining in traditional data sets, MapReduce architecture, Topic Modeling respectively. This  work attempts to integrate those three popular techniques in Big Twitter data stream to extract the valuable information hidden inside the large corpus and the benefits are multi-fold. Hence the architecture of proposed work is in Figure \ref{fig-ARCH} and the various modules are defined as follows:

\begin{figure}
\centering
\includegraphics[width=1.0\linewidth, height=0.3\textheight]{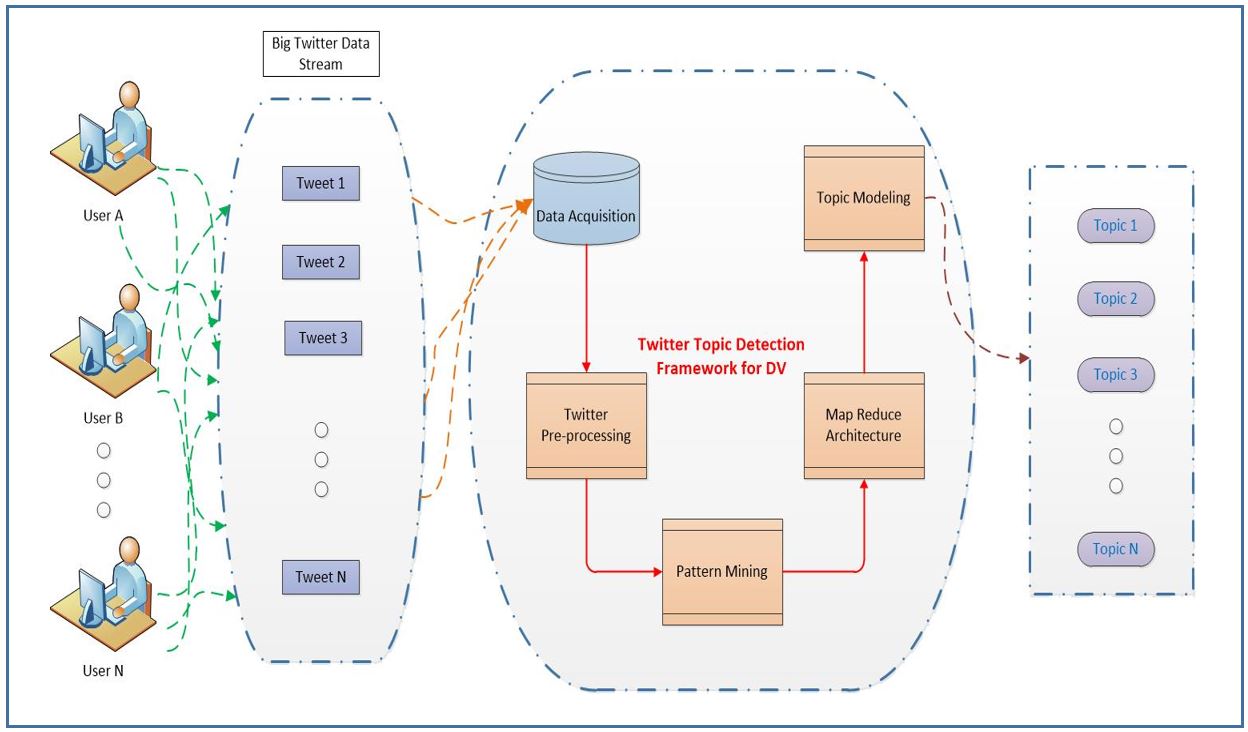}
\caption{Proposed Architecture for DV analysis}
\label{fig-ARCH}
\end{figure}

\begin{enumerate}
	\item  \emph{ Data Collection 
	\item  Data	Pre-processing
\item  Pattern Mining
	\item  MapReduce Framework
	\item  Topics prediction and Visualization
	\item  Evaluation Metrics}  
\end{enumerate}

The first stage is to collect DV related tweets from the social Media using Twitter API. Then, the pre-processing techniques are applied to remove noisy and inane content and extract the relevant data. From the pre-processed input data, semantic association between terms are captured by mining frequent patterns. 

In the next phase, the semantic units of the patterns are extracted and reduced using MapReduce architecture.  The frequent patterns generated from the twitter stream are taken as input to the mapper, which are shuffled and broad-casted to the reducer in the MapReduce architecture. This will output the terms that are more cohesively related to the topic. The purpose of this architecture is to extract the terms that are more relevant to the topic and the redundant terms are removed. In the earlier stage, we are mining the patterns, where the global terms are cohesively related to each other. Hence, this step augments the semantic worth to the conventional bag of words representation. 

The next stage models the process to generate the cluster of topics, which would be the summary of the entire Big Twitter Data stream. This step ensures that similar set of text words are clustered together and logically represents a theme for effective summarization. Then the cluster of topics is visualized using tag clouds. Finally, perform the evaluation metrics to predict the quality of topics.

 In the Data Collection process, the tweets are extracted from Twitter API based on keywords. In the second stage, the frequent patterns are generated that share many common terms related to abuse and mental disorders. \emph {$<$family violence, verbal abuse, depression$>$, $<$family violence, physical abuse, hurt, anxiety$>$, $<$domestic violence, sexual abuse, suicide$>$, and so on}. The Patterns exhibit similar semantic information leads to redundancy. Hence the Map reduce architecture is applied in the third category. It assigns over each pattern and Map iterates over each word in the pattern and finally produces the consolidated count of every word. In the next stage, the topic detection algorithm clusters the words based on the semantic structure like \emph{$<$physical abuse, verbal abuse, sexual abuse$>$} are clustered in to the topic \emph{``abuses"} and \emph{$<$depression, anxiety, suicide$>$} are clustered into the topic \emph{``mental disorders"} and so on. Thus this research produces actionable knowledge by converting the raw twitter data in to valuable information. This knowledge would be useful for the family health organizations, health care providers in further improving the health services, by monitoring the public social data. 

\subsection{Data Collection by Twitter Streaming API}
The constant stream of tweets arriving from the service is accessed through the Twitter Streaming API [50]. This module scrambles to collect as much data as possible, in order to find hidden patterns that can be acted upon for further knowledge discovery.  Apache Flume is a distributed and reliable data ingestion system that has a simple and flexible architecture for efficient collection of large amounts of Twitter data. The endpoints are configured in a workflow as “sources and sinks”. Every piece of data is defined as events in Flume, which are nothing but tweets. 
\begin{figure}
	\centering
		\includegraphics[width=.5\textwidth]{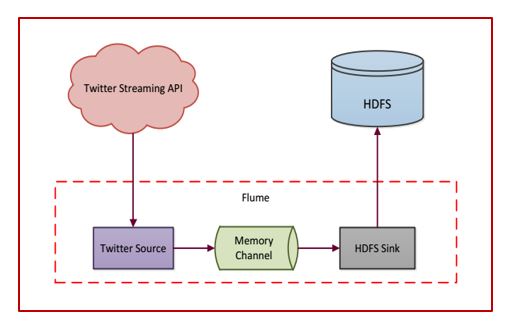}
	\caption{Flume architecture for the tweets collection}
	\label{flum}
\end{figure}

Figure \ref{flum} represents the flume architecture for high volume data gathering. The source generates events and where the data enters in to a flume. The acquired data flow through a channel and it is a drain between source and sink. Once the data reaches the sink, it writes the data to a Hadoop Distributed File System (HDFS). This module has to be designed a custom source to access the Twitter Streaming API, for the extraction of the relevant tweets based on search keywords and hashtags. It is worthless and time consumption to extract the complete Twitter firehose.  Initially we have to create four identical keys in our Twitter account: Consumer Key (API Key), Consumer Secret (API Secret), Access Token, and Access Token Secret and configure the flume source, channel and sink to gather the Twitter streaming data related to DV.  

\subsection{Pre-Processing of Twitter Stream}

The content of collected twitter stream varies from useful and meaningful information to incomprehensible text. The former has people opinion and relevant posts regarding the particular topic, whereas the latter may contain advertisements and non-worth reading.

Hence, in this step, high quality information and features are extracted by incorporating some pre-processing techniques. Pre-processing of Twitter stream removes noise that produces negatives effects and degrades performance. In microblogs, this is the most important step, as it contains the higher level of noise in the tweets. 
Stop word lists contain common English words like articles, prepositions, pronouns, etc., Examples are the, a, an, the, in, at, etc.  Hasan saif et al. [51] investigated that removing stop words improves the classification accuracy in Twitter analysis by reducing the data sparsity and shrinking the feature space. 

Stemming is used to identify the root of a word, to remove the suffixes related to a term, and to save memory space. For example, the words relations, related, relates all can be stemmed to relate.  Porter stemming [52][53] is applied to standardize terms appearance and to reduce data sparseness. Non text symbols and punctuation marks are removed.  Noisy tweets are filtered by eliminating links, non-ascii words, mentions, numbers and hashtags.

\subsection{Pattern Mining Model for Microblogs}
Mining patterns from Microblogs can resolve many problems mentioned in the previous section in the bag of words model representation. The unigrams are ambiguous and are interpreted differently with the words of different combinations. Hence, the algorithms have to be proposed to extract patterns in this module. This augments the relationship between the multiple terms relationship and co-occurrences of terms exhibit semantic associations.

Pattern mining discovers the recurring relationships and interesting correlations between terms. The quality of the detected topics depends on the quality of mined patterns. When extracted patterns need to satisfy the three qualities like frequency, collocation, completeness for higher accuracy and human interpretability. Frequency is the most important factor, how important is the frequency of pattern in the twitter stream, while detecting the topic. A collocation is referred as the co-occurrence of terms in the pattern in such a frequency that what is expected due to chance. Completeness is defined as the patterns provide the meaningful representation and has semantic meaning. The advantage of this step is as follows:
\begin{itemize}
\item \emph{Successful mining of patterns from tweets leads to discover high quality and informative topics.} 
\item \emph	{Pattern representation reduces noisy and redundant information and captures the words that are more meaningful to topics.}
\end{itemize}

\subsection{MapReduce Architecture}
MapReduce is a recent computing paradigm for distributed processing that process vast amount of data in parallel, reliable and fault tolerant manner. It has become much popular for data intensive programming model. A MapReduce job parallels the map tasks by decomposing the input data into separate chunks. The outputs of the maps are sorted out by giving as the input to the reduce class. Both the input and output of the MapReduce job are stored in a distributed file system.

In this step, it will consider frequent patterns consist of various terms as input and produces the appearances of each word wi that are hidden in the patterns. MapReduce framework is the most suitable method for pattern growth model, because of its functionality to split the patterns in to smaller portions. The two routines of MapReduce are Map and Reduce and they have different functionalities. 
A Map process is assigned to each pattern and Map iterates over each word wi in its assigned pattern and emits the pair $<w_i,1>$. Output from all equivalent classes of Map phases are grouped together and passed in to a single Reducer Phase. A serial algorithm computes the local task and the reducer finally counts the appearance of each word by gathering the output from all processes.  The purpose of this module is to generate the bag of frequent terms from the frequent patterns. The word has the highest count represents, its occurrence in most of the patterns and high contribution towards the topic. The words with the highest support count also signify its highest coherence in representing the semantic meaning in the patterns. 
\begin{itemize}
	\item \emph{The global frequent terms generated from the frequent patterns by MapReduce model, which are more relevant to topic.} 
	\item \emph	{Parallel execution of millions of patterns and hence increases the time efficiency.}
\end{itemize}

\subsection{Topic Prediction and Visualization}
In this module, the cluster of topics is generated for the global frequent terms, which are produced in the MapReduce architecture. Hence the various terms belongs to the each cluster topic depends on its semantic meaning and similarity. Hence, the corpus is the collection of topics and each topic is the distribution over terms. In this module, the domain knowledge in the form of Wikipedia or articles written about mental disorders, physical illness, sentiments and emotions and abuses related to DV are scrapped from the Web. The web crawlers extract the example of words related to various above mentioned themes in general. With this prior knowledge, a must-link technique models the extracted terms from MapReduce into various topics. A must-link defines that there exists semantic relationship between two terms t1 and t2; thus two terms belong to the same topic. They always appear together as topical terms in the same topic across many disciplines. Hence, the good topical words have the characteristics of better readability, worth description about the topic and high discrimination feature against neighbour topics.

The terms describing the every topic are ranked according to the support of a term. It can be defined as the term’s importance is directly proportional to the number of patterns contain it. Hence, the term frequency is calculated according to the number of times, it appears in the patterns. Based on the term support, the terms are ranked in the descending order. This importance ranking of all terms in all topics describes well the each of the detected topic in a better way.

The tag cloud [54] is an effective way to provide the summarization of terms describing the topics in a visually appealing way. The top frequent terms describing the topic are visualized using a tag cloud and the tags are usually terms. The tag cloud is useful to quickly perceiving the most prominent terms in the tweets and for differentiating the popular terms based on its font size and colour.

\subsection{Evaluation Metrics}

The topic detection performance can be evaluated from the result of classifying the texts to its relevant topic clusters. It can be measured by the following metrics: Precision, Recall, and F measure [55]. Precision and recall are the basic measures used to evaluate the performance of the proposed algorithm. True Positive (TP) denotes the number of terms correctly classified as relevant to the topic; False Positive (FP) denotes the number of irrelevant terms classified as relevant; True Negative (TN) denotes the number of terms correctly classified as irrelevant; False Negative (FN) denotes the number of terms misclassified as irrelevant. Thus, recall and precision are calculated as

\begin{equation}\label{eq:2}
Recall\left ( R \right ) =\frac{TP}{TP+ FN}
\end{equation}
\begin{equation}\label{eq:1}
Precision\left ( P \right ) =\frac{TP}{TP+ FP}
\end{equation}

Another metric F-Measure is also often used to measure the performance in information retrieval. It combines precision and recall with an equal weight in the following equation.

\begin{equation}\label{eq:3}
F-Measure = 2 \frac {PR}{P+R}
\end{equation}
In addition, the quality of the topics is detected by quantitative methods proposed by chang et al. [56] in the paper ``How Humans Interpret Topic Models” to evaluate the quality of the discovered topics". This can be predicted by two metrics: \emph{Term Intrusion} and \emph {Domain Expert Evaluation} to measure the semantic meaning in the topics inferred. 

Term intrusion assess whether the discovered topics have semantic coherence by the human perception. This task involves the bunch of questions to discover the intruder object from the options available. Each question contains 5 terms. 4 of them are randomly selected from top 10 terms of one topic and another term is an irrelevant one, which is chosen from another topic. The annotators have to analyse the terms and find out the incorrect term. The average of results that are answered correctly would be used to evaluate how well the terms are uniquely correlated and associated to each topic.

As this topic related to DV issue, we need the knowledge of domain expert. 
Once the topics like various abuses that women suffering from DV, their mental disorders, physical health problems, emotions and so on are discovered, we need the domain experts to analyse the various health problems that are extracted by our methodology is desirable. This can be evaluated by topic coherence property.  The terms in the topics are really coherent to topic thematic structure. 

\section{Functional Architecture}

\begin{figure}
	\centering
	\includegraphics[width=.5\textwidth]{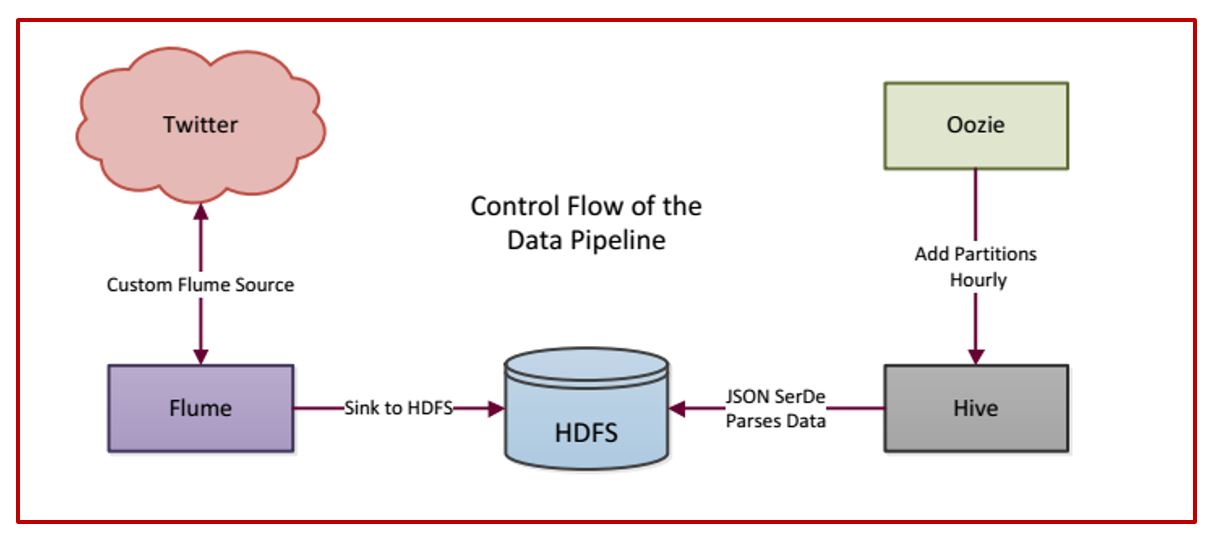}
	\caption{Hadoop Framework}
	\label{hadoop}
\end{figure}

Twitter has massive data storage and high processing requirements and thus it is difficult to handle with traditional databases. So, the paper implements the proposed methodological framework in Apache Hadoop platform [63] (Figure \ref{hadoop}) for the optimized data storage and workflow solutions and it is an open source distributed software platform.

\subsection{Gathering Twitter Data with Apache Flume}
As mentioned in previous section 4.1, the Twitter Data is extracted using the Apache Flume with Twitter Streaming API. Then the data can be stored in the Hadoop Distributed File System (HDFS).

\subsection{Storing data in HDFS}
The data can be stored in the Hadoop Distributed File System (HDFS), the prominent tool to store Big Twitter Data with high performance access. HDFS has the capability to support large scale applications with huge file sizes, where individual file even has the capacity of terabyte. HDFS mainly has two componets : Namenode and datanodes. This follows Master Slave architecture. There are multiple Datanodes, where the actual data resides and  a single Master Namenode, which monitors the information about data storage, size in the form of metadata. Resilience and fault tolerance are the key features of HDFS, where the data is divided into 64 MB chunks by default and replicated thrice across the data nodes.
\subsection{Querying Complex Data with Hive}
The Twitter Streaming API outputs tweets in a JSON format which are arbitrarily complex. Once the data loaded into HDFS, create the the external table to enable querying the data. Hence, Hive provides a query interface which can be used to query data that resides in HDFS in a simple way.The query language is similar to SQL, and  query the data by reading from HDFS. The interfaces Serializer and Deserializer (SerDe) tell the Hive to process the data and to flexibly define and redefine, by reading the JSON data and translates into objects for Hive.
\subsection{Partition Management with Oozie}
The Flume will continue to create new files, as the Twitter API perpetually stream the tweets.
When the new tweet arrives in, there is a need to add partitions to the Hive table periodically in an automatic way. Thus, Apache Oozie is a flexible workflow coordination system that instructs the job workflows to run based on a set of criteria (to refresh on a hourly basis).
\section{Conclusion}
This paper proposes a novel framework to extract actionable knowledge from social media discourse and this knowledge would be valuable to the the social welfare of the community.
This work will provide some significant results as it has the ``some unique advantage of real time access to naturalistic information by identifying the health related trends from the real world events and online communities" rather than traditional survey.

\section{References}
1.	S. Alhabib, U. Nur, and R. Jones, ``Domestic violence against women: Systematic review of prevalence studies," Journal of family violence, vol. 25, pp. 369-382, 2010.\\
2.	M. Ellsberg, H. A. Jansen, L. Heise, C. H. Watts, and C. Garcia-Moreno, ``Intimate partner violence and women's physical and mental health in the WHO multi-country study on women's health and domestic violence: an observational study," The Lancet, vol. 371, pp. 1165-1172, 2008.\\
3.	L. A. Goodman, M. P. Koss, and N. F. Russo, ``Violence against women: Physical and mental health effects. Part I: Research findings," Applied and Preventive Psychology, vol. 2, pp. 79-89, 1993.\\
4.	L. Barbosa and J. Feng, ``Robust sentiment detection on twitter from biased and noisy data," in Proceedings of the 23rd International Conference on Computational Linguistics: Posters, 2010, pp. 36-44.
5.	A.Tumasjan, T. O. Sprenger, P. G. Sandner, and I. M. Welpe, ``Predicting elections with twitter: What 140 characters reveal about political sentiment," ICWSM, vol. 10, pp. 178-185, 2010.\\
6.	B. O'Connor, R. Balasubramanyan, B. R. Routledge, and N. A. Smith, ``From Tweets to Polls: Linking Text Sentiment to Public Opinion Time Series," ICWSM, vol. 11, p. 1.2, 2010.\\
7.	T. Sakaki, M. Okazaki, and Y. Matsuo, ``Earthquake shakes Twitter users: real-time event detection by social sensors," in Proceedings of the 19th international conference on World wide web, 2010, pp. 851-860.\\
8.	R. Chunara, J. R. Andrews, and J. S. Brownstein, ``Social and news media enable estimation of epidemiological patterns early in the 2010 Haitian cholera outbreak," The American journal of tropical medicine and hygiene, vol. 86, pp. 39-45, 2012.\\
9.	S. Petrović, M. Osborne, and V. Lavrenko, ``Streaming first story detection with application to twitter," in Human Language Technologies: The 2010 Annual Conference of the North American Chapter of the Association for Computational Linguistics, 2010, pp. 181-189.\\
10.	http://www.internetlivestats.com/twitter-statistics/\\
11.	B. Pang and L. Lee, ``Opinion mining and sentiment analysis," Foundations and trends in information retrieval, vol. 2, pp. 1-135, 2008.\\
12.	B.Liu,``Opinion Mining",invited contribution to Encyclopedia of Database Systems,2008. \\
13.	J. Bollen, H. Mao, and A. Pepe, ``Modeling public mood and emotion: Twitter sentiment and socio-economic  phenomena," ICWSM, vol. 11, pp. 450-453, 2011.\\
14.	J. Bollen, H. Mao, and X. Zeng, ``Twitter mood predicts the stock market," Journal of Computational Science, vol. 2, pp. 1-8, 2011.\\
15.	A.Bruns and J. Burgess, ``\# ausvotes: How Twitter covered the 2010 Australian federal 
election," Communication,  Politics \& Culture, vol. 44, p. 37, 2011.\\
16.	A.Burns and B. Eltham, ``Twitter free Iran: An evaluation of Twitter's role in public diplomacy and information operations in Iran's 2009 election crisis," 2009.\\
17.	D. Gaffney, ``\# iranElection: quantifying online activism," 2010.\\
18.	A. Culotta, ``Detecting influenza outbreaks by analyzing Twitter messages," arXiv preprint arXiv:1007.4748, 2010.\\
19.	A. Culotta, ``Towards detecting influenza epidemics by analyzing Twitter messages," in Proceedings of the first workshop on social media analytics, 2010, pp. 115-122.\\
20.	E. de Quincey and P. Kostkova, ``Early warning and outbreak detection using social networking websites: The potential of twitter," in Electronic healthcare, ed: Springer, 2009, pp. 21-24.\\
21.	R. Chunara, J. R. Andrews, and J. S. Brownstein, ``Social and news media enable estimation of epidemiological patterns early in the 2010 Haitian cholera outbreak," The American journal of tropical medicine and hygiene, vol. 86, pp. 39-45, 2012.\\
22.	J. C. Bosley, N. W. Zhao, S. Hill, F. S. Shofer, D. A. Asch, L. B. Becker, et al., ``Decoding twitter: Surveillance and trends for cardiac arrest and resuscitation communication," Resuscitation, vol. 84, pp. 206-212, 2013.\\
23.	N. Heaivilin, B. Gerbert, J. Page, and J. Gibbs, ``Public health surveillance of dental pain via Twitter," Journal of dental research, vol. 90, pp. 1047-1051, 2011.\\
24.	A. Culotta, ``Lightweight methods to estimate influenza rates and alcohol sales volume from Twitter messages," Language resources and evaluation, vol. 47, pp. 217-238, 2013.\\
25.	N. K. Cobb, A. L. Graham, M. J. Byron, and D. B. Abrams, ``Online social networks and smoking cessation: a scientific research agenda," Journal of medical Internet research, vol. 13, p. e119, 2011.\\
26.	M. J. Paul and M. Dredze, ``Drug Extraction from the Web: Summarizing Drug Experiences with Multi-Dimensional Topic Models," in HLT-NAACL, 2013, pp. 168-178.\\
27.	S. A. Golder and M. W. Macy, ``Diurnal and seasonal mood vary with work, sleep, and daylength across diverse cultures," Science, vol. 333, pp. 1878-1881, 2011.\\
28.	M. Odlum and S. Yoon, ``What can we learn about the Ebola outbreak from tweets?," American journal of infection control, vol. 43, pp. 563-571, 2015.\\
29.	M. J. Paul and M. Dredze, ``A model for mining public health topics from Twitter," Health, vol. 11, pp. 16-6, 2012.\\
30.	M. J. Paul and M. Dredze, ``Discovering health topics in social media using topic models," PLoS One, vol. 9, p. e103408, 2014.\\
31.	M. J. Paul and M. Dredze, ``You are what you Tweet: Analyzing Twitter for public health," ICWSM, vol. 20, pp. 265-272, 2011.\\
32.	Q. He, K. Chang, and E.-P. Lim, ``Analyzing feature trajectories for event detection," in Proceedings of the 30th annual international ACM SIGIR conference on Research and development in information retrieval, 2007, pp. 207-214.\\
33.	J. Allan, ``Topic Detection and Tracking: Event Based Information Retrieval," Norvell, MA, USA: Kluwer Academic Publishers, 2002.\\
34.	G. Salton, ``Automatic text processing: The transformation, analysis, and retrieval of," Reading: Addison-Wesley, 1989.\\
35.	M. S. Bernstein, B. Suh, L. Hong, J. Chen, S. Kairam, and E. H. Chi, ``Eddi: interactive topic-based browsing of social status streams," in Proceedings of the 23nd annual ACM symposium on User interface software and technology, 2010, pp. 303-312.\\
36.	J. Sankaranarayanan, H. Samet, B. E. Teitler, M. D. Lieberman, and J. Sperling, ``Twitterstand: news in tweets," in Proceedings of the 17th acm sigspatial international conference on advances in geographic information systems, 2009, pp. 42-51.\\
37.	A. Marcus, M. S. Bernstein, O. Badar, D. R. Karger, S. Madden, and R. C. Miller, ``Twitinfo: aggregating and visualizing microblogs for event exploration," in Proceedings of the SIGCHI conference on Human factors in computing systems, 2011, pp. 227-236.\\
38.	C.-H. Lee, C.-H. Wu, and T.-F. Chien, ``BursT: a dynamic term weighting scheme for mining microblogging messages," in Advances in Neural Networks–ISNN 2011, ed: Springer, 2011, pp. 548-557.\\
39.	D. M. Blei, A. Y. Ng, and M. I. Jordan, ``Latent dirichlet allocation," the Journal of machine Learning research, vol. 3, pp. 993-1022, 2003.\\
40.	D. M. Blei and J. D. Lafferty, ``Dynamic topic models," in Proceedings of the 23rd international conference on Machine learning, 2006, pp. 113-120.\\
41.	L. AlSumait, D. Barbará, and C. Domeniconi, ``On-line lda: Adaptive topic models for mining text streams with applications to topic detection and tracking," in Data Mining, 2008. ICDM'08. Eighth IEEE International Conference on, 2008, pp. 3-12.\\
42.	D. Ramage, S. T. Dumais, and D. J. Liebling, ``Characterizing Microblogs with Topic Models," ICWSM, vol. 10, pp. 1-1, 2010.\\
43.	Agrawal R, Srikant R (1994) Fast algorithm for mining association rules in large database. In: Bocca JB, Jarke, M, Zaniolo C (eds.) VLDB’94. In: Proceeding of the 20th international conference on VLDB. pp 487–499, Santiago.\\
44.	R. Agrawal, T. Imieliński, and A. Swami, ``Mining association rules between sets of items in large databases," ACM SIGMOD Record, vol. 22, pp. 207-216, 1993.\\
45.	H.-G. Li, G.-Q. Wu, X.-G. Hu, J. Zhang, L. Li, and X. Wu, ``K-means clustering with bagging and mapreduce," in System Sciences (HICSS), 2011 44th Hawaii International Conference on, 2011, pp. 1-8.\\
46.	G. Zhang and M. Zhang, ``The Algorithm of Data Preprocessing in Web Log Mining Based on Cloud Computing," in 2012 International Conference on Information Technology and Management Science (ICITMS 2012) Proceedings, 2013, pp. 467-474.\\
47.	V. López, S. del Río, J. M. Benítez, and F. Herrera, ``Cost-sensitive linguistic fuzzy rule based classification systems under the MapReduce framework for imbalanced big data," Fuzzy Sets and Systems, vol. 258, pp. 5-38, 2015.\\
48.	M. Khan, ``Detecting Document Similarity in Large Document Collection using MapReduce and the Hadoop Framework," Brac University, 2012.\\
49.	F. Chen and M. Hsu, ``A performance comparison of parallel DBMSs and MapReduce on large-scale text analytics," in Proceedings of the 16th International Conference on Extending Database Technology, 2013, pp. 613-624.\\
50.	https://dev.twitter.com/streaming/overview\\
51.	H. Saif, M. Fernández, Y. He, and H. Alani, ``On stopwords, filtering and data sparsity for sentiment analysis of Twitter," 2014.
52.	M. F. Porter, ``Snowball: A language for stemming algorithms," ed, 2001.\\
53.	M. F. Porter, ``An algorithm for suffix stripping," Program, vol. 14, pp. 130-137, 1980.\\
54.	https://en.wikipedia.org/wiki/Tag\_cloud \\
55.	D. M. Powers, ``Evaluation: from precision, recall and F-measure to ROC, informedness, markedness and correlation," 2011.\\
56.	J. Chang, S. Gerrish, C. Wang, J. L. Boyd-Graber, and D. M. Blei, ``Reading tea leaves: How humans interpret topic models," in Advances in neural information processing systems, 2009, pp. 288-296.\\
57.	Z. He, X. Xu, and S. Deng, ``Data mining for actionable knowledge: A survey," 2005.\\
58.	C. Ling, T. Chen, Q. Yang, and J. Chen, ``Mining optimal actions for intelligent CRM," in Proc. IEEE Int’l Conf. Data Mining (ICDM), 2002.\\
59.	M. Mathioudakis and N. Koudas, ``Twittermonitor: trend detection over the twitter stream," in Proceedings of the 2010 ACM SIGMOD International Conference on Management of data, 2010, pp. 1155-1158.\\
60.	J. Weng and B.-S. Lee, ``Event Detection in Twitter," ICWSM, vol. 11, pp. 401-408, 2011.\\
61.	G. Petkos, S. Papadopoulos, L. Aiello, R. Skraba, and Y. Kompatsiaris, ``A soft frequent pattern mining approach for textual topic detection," in Proceedings of the 4th International Conference on Web Intelligence, Mining and Semantics (WIMS14), 2014, p. 25.\\
62.	J. Huang, M. Peng, and H. Wang, ``Topic Detection from Large Scale of Microblog Stream with High Utility Pattern Clustering," in Proceedings of the 8th Workshop on Ph. D. Workshop in Information and Knowledge Management, 2015, pp. 3-10.\\
63.	W. Tom. ``Hadoop: The definitive guide",O'Reilly Media, Inc, 2012.\\
64. L. Sun, H. Wang, J. Soar, and C. Rong, ``Purpose based access control
for privacy protection in e-healthcare services," Journal of Software,
vol. 7, no. 11, pp. 2443-2449, 2012. \\
65. L. Sun, H.Wang, J. Yong, and G.Wu, ``Semantic access control for cloud
computing based on e-healthcare," in Computer Supported Cooperative
Work in Design (CSCWD), 2012 IEEE 16th International Conference
on. IEEE, 2012, pp. 512-518. \\
66. H. Wang, Z. Zhang, and T. Taleb, ``Special issue on security and privacy
of iot," World Wide Web, pp. 1-6, 2017. \\
67. X. Sun, H. Wang, J. Li, and Y. Zhang, ``Satisfying privacy requirements
before data anonymization," The Computer Journal, vol. 55, no. 4, pp.
422-437, 2012. \\
68. J. Li, H. Wang, H. Jin, and J. Yong, ``Current developments of kanonymous
data releasing," Electronic Journal of Health Informatics,
vol. 3, no. 1, p. 6, 2008. \\
69. H. Wang, J. Cao, and Y. Zhang, ``A flexible payment scheme and its
role-based access control," IEEE Transactions on knowledge and Data
Engineering, vol. 17, no. 3, pp. 425-436, 2005. \\
70. H. Wang and L. Sun, ``Trust-involved access control in collaborative
open social networks," in Network and System Security (NSS), 2010 4th
International Conference on. IEEE, 2010, pp. 239-246. \\
71. J. Huang, M. Peng, H. Wang, J. Cao, W. Gao, and X. Zhang, ``A
probabilistic method for emerging topic tracking in microblog stream,"
World Wide Web, vol. 20, no. 2, pp. 325-350, 2017. \\
72. H. Wang, Y. Zhang et al., ``Detection of motor imagery eeg signals
employing na¨ıve bayes based learning process," Measurement, vol. 86,
pp. 148-158, 2016. \\
73. H. Hu, J. Li, H. Wang, G. Daggard, and M. Shi, ``A maximally
diversified multiple decision tree algorithm for microarray data classification,"
in Proceedings of the 2006 workshop on Intelligent systems
for bioinformatics-Volume 73. Australian Computer Society, Inc., 2006,
pp. 35-38. \\
74. Y. Wang, H. Li, H. Wang, B. Zhou, and Y. Zhang, ``Multi-window based
ensemble learning for classification of imbalanced streaming data,"
in International Conference on Web Information Systems Engineering.
Springer, 2015, pp. 78-92. \\
75. X. Yi and Y. Zhang, ``Privacy-preserving naive bayes classification on
distributed data via semi-trusted mixers," Information systems, vol. 34,
no. 3, pp. 371-380, 2009. \\



\end{document}